\documentstyle[12pt,epsfig]{article}
\begin{document}

\begin{titlepage}
\rightline{December 2008}

\vskip 2.5cm
\centerline{\Large \bf Dilaton as the Higgs boson}
\vskip 2cm
\centerline{Robert Foot$^{a}$, Archil Kobakhidze$^{a}$, Kristian L. McDonald$^{b}$\footnote{Electronic address:
rfoot@unimelb.edu.au, archilk@unimelb.edu.au, klmcd@triumf.ca}}
\vskip 1cm
\centerline{$^{a}$ \ School of Physics,}
\vskip 0.2cm
\centerline{\ The University of Melbourne, Victoria 3010,
Australia}
\vskip 0.4cm
\centerline{$^{b}$ \ Theory Group, TRIUMF, 4004 Wesbrook Mall,
Vancouver, BC V6T 2A3, Canada}

\vskip 4cm
\noindent
We propose a model where the role of the electroweak Higgs field is played by the 
dilaton. The model contains terms which explicitly violate 
gauge invariance, however it is shown that this violation is fictitious, so that the model is a consistent 
low energy effective theory. 
In the simplest version of the idea the resulting low energy effective
theory is the same as the top mode standard model.

\end{titlepage}

\newpage

\baselineskip=16pt

\paragraph{Introduction}

Fundamental vector fields are the carriers of the strong, weak and electromagnetic interactions 
of the Standard Model. A consistent quantum mechanical description of these vector 
fields (massless and massive) is possible within the context of locally gauge invariant theories. 
More specifically the massive weak bosons $W^{\pm}_{\mu}$ and $Z_{\mu}$ are treated as the gauge 
bosons of a spontaneously broken $SU(2)\times U(1)_Y$ electroweak gauge 
symmetry~\cite{Glashow:1961}, which is simply realized through the Higgs 
mechanism~\cite{Higgs:1964ia}. 
The theory predicts the existence of an electrically neutral scalar field called the 
Higgs boson. The Higgs boson interacts with other Standard Model particles, 
notably with the weak gauge bosons, with couplings dictated by
gauge invariance. As a result, scattering amplitudes of the 
longitudinal modes of massive weak bosons satisfy the unitarity bound, 
provided the Higgs boson is not too heavy, $m_h< 1$ TeV or so. 

The dilaton field is motivated in many extensions of the Standard Model which attempt to 
consistently incorporate gravitational interactions. In fact many known extensions of the 
Einstein theory of gravitation are of the Brans-Dicke type scalar-tensor theories. 
For example, in string theory the graviton is inevitably accompanied by the dilaton, 
whilst in models of electroweak symmetry breaking with a near or exact 
scale invariance the dilaton may play a role in low energy phenomenology (see e.g., \cite{Foot:2007iy}, \cite{Goldberger:2007zk}). 
If the dilaton field is indeed present in the particle spectrum, one may ask whether 
it can be identified with the Higgs boson. This is the question we would like to discuss in 
the present paper\footnote{See also~\cite{Chernodub:2008rz} for another 
interesting non-standard analysis of the electroweak symmetry breaking sector.}.   

At first glance it appears that the dilaton can not play the role of the Higgs boson since 
it does not couple to the Standard Model fields in a gauge invariant way. 
Actually, because the Standard Model Lagrangian without the Higgs sector is 
classically scale-invariant, the dilaton can be completely removed at the classical level 
by an appropriate conformal transformation. Therefore the scale-invariance of the Standard Model 
must be broken in order to couple the dilaton to matter fields. In this work we consider models 
where scale invariance is broken explicitly by mass terms for the fermion fields. 
These fermion mass terms also explicitly break gauge invariance and thus it seems we achieve nothing. 
Nevertheless, we show that if the dilaton is a non-dynamical degree of freedom classically, 
the above explicit breaking of gauge invariance is actually fictitious. The key point is that the system 
possesses a gauge invariant constraint, induced by the equation of motion for 
the non-dynamical dilaton field, which enforces full gauge invariance.
It turns out that the simplest implementation of this idea reduces to the gauged Nambu-Jona-Lasinio model \cite{Nambu:1961tp}, 
which allows for dynamical electroweak symmetry breaking where the dilaton plays the role of the Higgs boson.

\paragraph{The basic idea}

Let us start by considering a locally gauge invariant theory with gauge group $G$. 
Besides the gauge fields, $A_{\mu}\equiv A_{\mu}^{a}T^{a}$ ($T^a$ are generators of $G$), we introduce two 
fermion fields $\psi(x)$ and $\chi(x)$, as well as a non-dynamical real scalar field $h(x)$. 
These fields transform under $G$ as
\begin{eqnarray}
G:~~ A_{\mu}\to U A_{\mu} U^{-1}+\frac{i}{g}U\partial_{\mu}U^{-1}~,~~
\psi \to U\psi~,~~\chi \to \chi~~{\rm and}~~h\to h~,
\label{a}
\end{eqnarray}
where $U(x) \in G$. The Lagrangian of the theory is written as
\begin{eqnarray}
{\cal L} = -\frac{1}{2}{\rm Tr}(F_{\mu\nu}F^{\mu\nu}) + i\bar \psi \gamma^{\mu}\left(D_{\mu} \psi\right)
+ i\bar \chi \gamma^{\mu}\left(\partial_{\mu} \chi\right) +\left[\lambda \bar \psi h \chi + {\rm h.c.}\right ],
\label{b}
\end{eqnarray}
where $D_{\mu}\equiv \partial_{\mu}-igA_{\mu}$ is the covariant
derivative and $F_{\mu\nu}=\frac{i}{g}[D_{\mu}, D_{\nu}]$ is the
$G$-valued field strength. Except for the Yukawa interaction terms in
parenthesis, the above Lagrangian is invariant under the local gauge
transformation (\ref{a}). Note also that the kinetic term for $h$ is
absent in (\ref{b}), i.e. $h$ is non-dynamical. Because of these apparent
drawbacks, the model described by the Lagrangian (\ref{b}) can not be
renormalizable. We assume that it holds at a certain high energy scale
$\Lambda$. Moreover, since the gauge 
invariance is broken \emph{explicitly} one might expect violation of unitarity in, e.g., 
the scattering amplitudes of longitudinal gauge field modes. The latter expectation is wrong however. 
It turns out that, due to the non-dynamical nature of the scalar field $h$, the explicit breaking of 
gauge invariance is fictitious in the sense that the Lagrangian (\ref{b}) actually describes 
gauge invariant dynamics. To see this we rewrite (\ref{b}) in an equivalent form with manifest gauge invariance.

The fact that (\ref{b}) describes gauge invariant dynamics is far from obvious. Indeed with the 
field variables used in (\ref{b}) it is impossible to solve in closed form  all the constraints  
which enforce the gauge invariant dynamics. Fortunately, we can introduce new field variables such that 
the gauge invariant constraints can be solved explicitly. To this end, let us consider the 
unimodular scalar field $\Phi \in G$, $\Phi^{\dagger}\Phi=1$, which transforms as $\Phi\to U\Phi$ under $G$, 
and define new field variables:
\begin{equation}
h\to H = \Phi^{\dagger} h~,~~\psi \to \psi'=\Phi^{\dagger} \psi~,~~A_{\mu}\to A'_{\mu}= 
\Phi^{\dagger}\left( A_{\mu}+\frac{i}{g}\partial_{\mu}\right )\Phi~ .\label{b2}  
\end{equation}    
In terms of these fields the Lagrangian (\ref{b}) takes the form, 
\begin{eqnarray}
{\cal L} = -\frac{1}{2}{\rm Tr}(F_{\mu\nu}'F'^{\mu\nu}) + i\bar \psi' \gamma^{\mu}\left(D'_{\mu} \psi'\right)
+ i\bar \chi \gamma^{\mu}\left(\partial_{\mu} \chi\right) +\left[\lambda \bar \psi' H \chi + {\rm h.c.}\right ].
\label{c}
\end{eqnarray} 
It is important to note that (\ref{c}) is not invariant under  $G$. The field $H$ is again non-dynamical, 
and can be viewed as a Lagrange multiplier which imposes the \emph{gauge invariant} constraint
\begin{equation}
\bar \psi'  \chi = 0~.
\label{d}
\end{equation}
Implementing this constraint explicitly in the 
functional integral\footnote{This is equivalent to integrating out the auxiliary field $H$ in the functional integral.} 
gives the equivalent Lagrangian
\begin{eqnarray}
{\cal L}_{\mathrm{equiv.}} = -\frac{1}{2}{\rm Tr}(F_{\mu\nu}F^{\mu\nu}) + i\bar \psi \gamma^{\mu}\left(D_{\mu} \psi \right)
+ i\bar \chi \gamma^{\mu}\left(\partial_{\mu} \chi\right) +
\frac{\lambda^2}{\mu^2}
\left(\bar \psi \chi \right)
\left(\bar \chi \psi\right),
\label{e}
\end{eqnarray} 
where\footnote{We note by passing that the Lagrangian (\ref{e}) describes the four-fermion  
Nambu--Jona-Lasinio model~\cite{Nambu:1961tp} with the four-fermion coupling 
$G_{\mathrm{4f}}=\frac{\lambda^2}{\mu^2}$. The NJL model is often used in phenomenological studies of 
dynamical symmetry breaking.} $\mu^2\to 0$. This Lagrangian is equivalent to (\ref{c}) and is 
manifestly gauge invariant under the transformations (\ref{a}). Note that in the above manipulations 
no gauge fixing of the group $G$ is involved; we merely changed the field variables which does 
not affect the symmetry properties of the functional integral measure.  Therefore, the claim that 
the system is gauge invariant is an exact statement, i.e. it is true for the full quantum theory, 
not just for the classical one. 

Observe that in going from equation (\ref{b}) 
to equation (\ref{c}) the correct implementation of the parameterization (\ref{b2}) does not introduce 
any new degrees of freedom. To see this more explicitly, we note that after the 
change of variables (3), the Lagrangian (4) is formally invariant under a new 
(different from the initial $G$) $G'$ local gauge transformations:
\begin{equation}
G':~~ A'_{\mu}\to U' A_{\mu} U'^{-1}+\frac{i}{g}U'\partial_{\mu}U'^{-1}~,~~
\psi' \to U'\psi'~,~~\chi \to \chi~~{\rm and}~~H\to  U' H~,
\label{f}
\end{equation}
where $U'(x)\in G'$. The $G'$ gauge freedom ensures that the number of degrees of freedom described by 
the new variables $(A'_{\mu}, \Phi)$ is exactly equal to the number of degrees of freedom 
described by the old field variables $A_{\mu}$. The most illuminating choice is to take the Landau gauge, i.e.
\begin{equation}
\partial^{\mu}A_{\mu}'^a=0~,
\label{g}
\end{equation}
where $a=1,2,...n=\mathrm{dim}(G')$.
Note that this is a gauge fixing condition  with respect to the $G'$ gauge symmetry, 
but merely a \emph{gauge invariant} transversality condition on the vector field $A_{\mu}'$ with 
respect to the initial $G$ gauge symmetry. That is to say, the $\Phi (x)$ field describes $n$ longitudinal 
degrees of freedom of a vector field $A_{\mu}(x)$, not new degrees of 
freedom\footnote{For an Abelian symmetry the condition (\ref{g}) can be solved in closed form to 
express $\Phi$ through $A_{\mu}$, $\log(\Phi)=ig\frac{\partial^{\mu}}{\Box}A_{\mu}$.}. Therefore the 
lagrangians (\ref{b}) and (\ref{c}) are indeed fully equivalent.

Summarizing the lesson we have learned, in theories with explicit gauge
breaking terms certain gauge invariant constraints might enforce the
full gauge invariance. The degrees of freedom ``missing" in the
unconstrained gauge non-invariant theory must be non-dynamical at the
classical level in order to generate the desired constraints through
their equations of motion. These degrees of freedom become dynamical at
the quantum level. Based on this observation, we now present a realistic
model of electroweak symmetry breaking where the role of the Higgs boson
is played by the dilaton field.

\paragraph{The model}

We would like to describe a scenario where the role of the Higgs boson is played by the dilaton, 
which, as was discussed above, must be non-dynamical at tree level. This is the case when the dilaton 
field $\phi(x)$ couples to gravity with the conformal coupling $\xi=1/6$, 
\begin{equation}
{\cal L}_{\mathrm{grav}}=
\sqrt{-\hat g}\left[-\frac{\xi}{2}\phi^2\hat R+
\frac{1}{2}\hat g^{\mu\nu}\partial_{\mu}\phi\partial_{\nu}\phi-\frac{1}{2}\mu^2 \phi^2\right]~.
\label{m1}
\end{equation}
In the above Lagrangian we have also included the dilaton mass term. This mass term explicitly violates the 
classical scale invariance of (\ref{m1}). We will have further comments on this Lagrangian later on.

Matter couples to gravity in the usual way through the diffeomorphism invariant Lagrangian, 
\begin{equation}
{\cal L}_{\mathrm{matter}}=\sqrt{-\hat g}{\cal L}_{\mathrm{SM}}(\hat g_{\mu\nu}, \hat F)+\sqrt{-\hat g}\left[\overline{\hat Q}_{L}\hat M \hat t_{\mathrm{R}}+\mathrm{h.c.}\right]~. 
\label{m2}
\end{equation}
Here $\cal{L}_{\mathrm{SM}}$ denote a diffeomorphism covariant form of the  Standard Model Lagrangian 
which involves all the Standard model fields, collectively denoted by $\hat F$, except the electroweak 
Higgs doublet. $\cal{L}_{\mathrm{SM}}$ is invariant under the $SU(3)\times SU(2)\times U(1)_Y$ local gauge 
transformations, as well as under the classical scale transformations. We add $SU(2)\times U(1)_Y$ violating 
fermionic mass terms, which also \emph{explicitly} violate the scale invariance. For the sake of simplicity, 
we consider only the dominant\footnote{In view of neutrino see-saw masses, the neutrino Dirac mass terms 
might actually be dominant. See the comment below.} top-quark mass 
term in (\ref{m2}), $\hat M=(m_{\mathrm{t}},0)^{\mathrm{T}}$, where $m_{\mathrm{t}}$ is some mass parameter 
and $\hat Q_{L}=(\hat t_L,\hat b_L)^{\mathrm{T}}$ is the quark doublet. We regard the Lagrangian 
${\cal L}={\cal L}_{\mathrm{grav}}+{\cal L}_{\mathrm{matter}}$ as a low-energy effective Lagrangian 
(in the Wilsonian sense) valid below some high energy scale $\Lambda$.

The metric $\hat{g}_{\mu\nu}$ is taken in the so-called Jordan frame.
From the particle physics standpoint however the Einstein frame must be
regarded as the physical one\footnote{The reason being that in the
Jordan frame the dilaton has a kinetic mixing term with the scalar (trace) part of the graviton field, 
while in the Einstein frame provides diagonal basis for the kinetic terms.}. We go to the Einstein frame by 
performing a Weyl rescaling of the the metric $g_{\mu\nu}=\Omega^2(x)\hat g_{\mu\nu}$, and fields, 
$F=\Omega^{S_F}(x)\hat F$ ($S_F$ is the conformal weight of the field $F$: $0$ for vector bosons, $-1$ for 
scalar bosons and $-3/2$ for fermions), taking $\Omega(x)$ such that the rescaled dilaton field is equal 
to the reduced Plank mass $M_{\mathrm{P}}=1/8\pi G_N \approx 2\cdot10^{18}$ GeV, i.e. $\Omega(x)=\phi(x)/M_{\mathrm{P}}$. 
As a result of these rescalings the dilaton field is removed from the scale invariant part of the Lagrangian and 
appears only in the scale noninvariant mass terms. Thus we have:  

\begin{equation}
\frac{1}{\sqrt{-g}}{\cal L}=-\frac{M^2_{\mathrm{P}}}{2} R+{\cal L}_{\mathrm{SM}}(g_{\mu\nu}, F)-\mu^2h^{\mathrm{T}}h
+\left[y_{\mathrm{top}}\overline{Q}_{L}h t_{\mathrm{R}}+\mathrm{h.c.}\right]~,
\label{m3}
\end{equation}     
where $h = (\frac{M_P^2}{\sqrt{2}\phi(x)},0)^{\mathrm{T}}$ and
$y_{\mathrm{top}}=\sqrt{2}m_{\mathrm{t}}/M_{\mathrm{P}}$. The Lagrangian (\ref{m3}) is phenomenologically equivalent 
to the fully 
gauge invariant NJL-type Lagrangian for the top 
mode Standard Model~\cite{Bardeen:1989ds}. To observe this note that structurally (\ref{m3}) is the same 
as the Lagrangian in (\ref{b}) and thus one may repeat the method above described. 
The classical equation of motion for $h$ will enforce a constraint such that after integrating out $h$ the 
full gauge invariance of the Lagrangian is manifest. The resulting form of the Lagrangian is equivalent to the 
NJL-type Lagrangian for the top 
mode Standard Model and thus describes the same phenomenology\footnote{Also, the conformal 
coupling $\xi=1/6$ in eq. (\ref{m1}) can be justified as it appears as a stable infrared renormalization group 
fixed point in such type of models \cite{Hill:1991jc}.}.

The top 
mode Standard Model is known to suffer from some phenomenological problems; for
example it prefers the top quark to be heavier than is experimentally observed.
However simple variations of our proposed minimal model are possible. 
For example, instead of the top quark mass one could use a Dirac neutrino mass term $m_D \bar \nu_L \nu_R$
as the dominate electroweak symmetry breaking mass scale (c.f. ref.\cite{ant:2003}), and include the term $\lambda \phi \bar \nu_R (\nu_R)^c$.
The resulting model would then be phenomenologically consistent and 
have the advantage of accommodating small neutrino masses via the see-saw mechanism. 
This modification would not alter our main point.

In conclusion, we have proposed an explicit model which enables the 
role of the Higgs boson to be played by the dilaton field. In the
simplest version the resulting low energy effective theory is 
equivalent to the top mode standard model. Simple variations,
accommodating neutrino masses, are also possible.

 \vskip 1cm
 \noindent
 {\bf Acknowledgements:} 
 This work was supported by the Australian Research Council.

\end{document}